\documentclass[english]{article}
\usepackage[T1]{fontenc}
\usepackage[latin9]{inputenc}
\usepackage{amssymb}
\usepackage{graphicx}
\usepackage{amsmath}
\usepackage{babel}
\providecommand{\keywords}[1]{\textbf{\textit{Keywords:}} #1}

\usepackage{amsfonts}
\usepackage[linesnumbered,ruled]{algorithm2e}



\usepackage{babel}
\begin{document}

\title{DiracSolver: a tool for solving the Dirac Equation}

\author{Ioannis G. Tsoulos$^{(1)}$, O.T. Kosmas$^{(2)}$, V.N. Stavrou$^{(3)}$}

\date{$^{(1)}$Department of Communications, Informatics and Management,
Technological Educational Institute of Epirus, Greece\\
 $^{(2)}$Modeling and Simulation Centre, MACE, University of Manchester, M13 9PL Manchester, UK\\
$^{(3)}$Military Institutes of University Education (ASEI), Hellenic Naval Academy, 18539 Piraeus, Greece}

\maketitle

\begin{abstract}
Advantageous numerical methods for solving the Dirac equations are 
derived. They are based on different stochastic optimization techniques,
namely the Genetic algorithms, the Particle Swarm Optimization and the 
Simulated Annealing method, their use of which is favored from
intuitive, practical, and theoretical arguments. Towards this end, 
we optimize appropriate parametric expressions representing the radial 
Dirac wave functions by employing methods that minimize multi 
parametric expressions in several physical applications. As a concrete 
application, we calculate the small (bottom) and large (top) components 
of the Dirac wave function for a  bound muon orbiting around a 
very heavy (complex) nuclear system (the $^{208}$Pb nucleus), but the 
new approach may effectively be applied in other complex atomic, 
nuclear and molecular systems.
\end{abstract}



\section{Introduction }

In many modern physical problems and specifically those related to the
structure and evolution of quantum systems, the evaluation of exact 
predictions for various physical observables is required 
\cite{Pap-Lag-1,Pap-Lag-2,Stoica-PRC,Stoica-AHEP}. 
These predictions are based on wave functions describing 
the states of the quantum system in question that 
come out of accurate solutions of specific type differential equations, 
like the Schroedinger and Dirac equations, etc. 
\cite{Stoica-PRC,Stoica-AHEP,Kosm-Lag}. 
Because, usually, analytic solutions of such differential equations
describing many body quantum systems are not possible, the application 
of advanced algorithms is required \cite{Kosm-Lag,Kitano}. 

Concrete examples of the category of physical problems we deal with in the 
present work are systems of leptons (electron, muon, tau particle, etc.) 
bound in extended Coulomb fields created by the protons of atomic nuclei 
\cite{Ziner,Giannak-Kosm}. It is well known that, such problems may be solved 
analytically in special cases and only under the assumption of point-like 
constituents of the systems, i.e. point-like nucleons, nuclei etc. (a fairly 
crude approximation), but the consideration of the finite size of 
nucleons and nuclei necessitates the use of advantageous numerical 
techniques \cite{Stoica-AHEP,Kosm-Lag,Kitano}. In the first case, 
the wave functions of e.g. a hydrogenic-type lepton-nucleus system are obtained, 
and the Schroedinger as well as the Dirac equations may be solved analytically. 
In the second case, however, for solving these equations, the application of 
modern numerical methods should be adopted in complete analogy with the 2-body 
problem where, after removing (or neglecting) the center of mass motion, one rewrites 
these equations in the relative lepton-nucleus coordinate system (assuming spherical 
symmetry of the Coulomb field, the differential equations to be solved, are 
expressed in terms of the relative coordinate $r$). Then, the main problem is 
restricted to solving the radial part of the above mentioned 3-Dimensional 
differential (Schroedinger or Dirac) equations \cite{Stoica-AHEP,Kosm-Lag,Kitano}.

In recent years, there is intense interest in treating numerical solutions of wave functions for many body problem through the use of the field of machine learning methods \cite{Science,Machine1,Machine2,Machine3}. 
In this work, we make an attempt to derive methods for solving the Dirac differential 
equations on the basis of the assumptions of stochastic methods. More specifically, 
we choose the genetic algorithms \cite{genetic}, the particle swarm optimization (PSO) 
\cite{kennedy} and the simulated annealing  method \cite{simulated} in the context of 
which, one may optimize appropriate multi-parametric expressions representing the 
radial part of the wave functions describing a lepton orbiting around a complex nucleus 
\cite{Pap-Lag-1,Pap-Lag-2,Kosm-Lag}.
 
It is worth noting that, within the context of these methods compact expressions 
for the radial part of the Schroedinger and Dirac wave functions, in the form of 
a rather simple summation times an exponential, both functions of the radial 
coordinate $r$, are obtained, which afterwards facilitate several physical applications.

In this paper, we applied the algorithms to obtain the ground state wave function describing 
a muon-nucleus system (muonic $^{208}$Pb atom), however, its application in order to find 
the wave functions for other quantum states like those of a tau mesic atom \cite{Oset-91}
or hydrogenic-type electron nucleus systems, etc.~\cite{Kitano}, is straightforward. 
From a computational point of view, the training in our method is performed by using the
DiracSolver software package that proved to be both convenient and efficient. This offers 
the possibility for the developed in this work new approach, to be effectively applied in 
other complex atomic, nuclear and molecular systems. 

The rest of the paper is organized as follows. In Section 2, the relativistic mathematical 
formulation and the construction of the error function of the problem is described. Then, the new 
minimization methods are presented (in Section 3) and the Software documentation is discussed 
(Section 4). These methods are applied in concrete physical systems and the obtained results for example 
runs are also presented and discussed in Section 5. Finally (Section 6) the main conclusions 
extracted in this work are summarized.

\section{Formulation of the mathematical problem }

In this section we concentrate on the description, with the sufficient needed detail,
of the theoretical formalism of the Dirac equations that refer to the motion of a 
quantum mechanical system moving in a central force field with relativistic velocity. 
The 3-Dimensional Dirac equations describing a Dirac particle in a central field are 
explained in the Appendix where the main ingredients and the required physical quantities 
are defined \cite{Rose}. In the next subsection, we focus on the analysis of the radial 
part of the coupled first order (ordinary) differential Dirac equations.

\subsection{The coupled radial Dirac equations}

In order to figure out the mathematical problem we face in this paper, we start with 
the description of the quantum system we deal with, that consists of a lepton of mass 
$m_\ell$ ($\ell = e, \mu, \tau$, for electron, muon, tau particle, respectively) 
bound in the electrostatic field created by an atomic nucleus. The latter contains
a number of $A$ nucleons ($Z$ of them are protons with mass $m_p$ and $N$ are 
neutrons with mass $m_n$, so as $A=Z+N$) having total mass $M=Nm_n + Zm_p - B(Z,N)$,
where $B(Z,N)$ denotes the binding energy of the nuclear system \cite{Feschach} 
(for convenience, B(Z,N) may initially be neglected). 
The reduced mass, $m$, of the bound, in such a system, lepton is given by
\begin{equation}
m = \frac{m_\ell}{1 + m_\ell/M}.
\label{eq:red-mass}
\end{equation}

In our investigation, we consider the two reduced radial wave functions of the 
Dirac equation for central force fields (see Appendix), i.e. the large (top), 
$f(r)$, and the small (bottom), $g(r)$, components that satisfy the differential 
equations (see for example \cite{Ziner,Kitano})
\begin{equation}
\label{1_Dir_Eq_fr}
\frac{d}{dr}f(r)+\frac{1}{r}f(r)=\frac{1}{\hbar c}(m c^{2}
-E+V(r))g(r) 
\end{equation}
\begin{equation}
\label{2_Dir_Eq_gr}
\frac{d}{dr}g(r)-\frac{1}{r}g(r)=\frac{1}{\hbar c}(m c^{2}
+E-V(r))f(r).
\end{equation}
In this work, for simplicity we restrict ourselves in the ground state (1s state) 
of the bound lepton-nucleus system and the quantum number $\kappa$ (see Appendix)
takes the value $\kappa=-1$ \cite{Kitano}. 
In Eqs. (\ref{1_Dir_Eq_fr}) and (\ref{2_Dir_Eq_gr}), $E$ denotes the total energy 
of the lepton, $c$ is the speed of light and ${\hbar c} = 197.327$ MeV fm.

For the description of the central force nuclear field, as mentioned in the
Introduction, we take into account the finite size of the studied nucleus 
by deriving the potential $V (r)$ from a realistic charge density distribution 
of the atomic nucleus (see Sect. 5). 

In order to deduce the error function to be minimized, we use the following 
parameterized solutions for the small and large components of the Dirac wave
function 
\begin{equation}
f(r)=re^{-\beta r}N\left(r,{\bf u_{f}},{\bf v_{f}},{\bf w_{f}}\right)\label{eq:frequation}
\end{equation}
and
\begin{equation}
g(r)=re^{-\beta r}N\left(r,{\bf u_{g}},{\bf v_{g}},{\bf w_{g}}\right)\label{eq:grequation}
\end{equation}
with $\beta>0$. The parameters ${\bf u_{g}}$, ${\bf v_{g}}$, ${\bf w_{g}}$ may
be assumed to be the parameters of a feed-forward artificial neural network 
$N\left(r,{\bf u_{g}},{\bf v_{g}},{\bf w_{g}}\right)$ \cite{Kosm-Lag,Bishop}. From the aforementioned parameters $\beta$ is more or less related to the wave number $k=\sqrt(2 m_\ell E)$ of the free particle wave function. $u_{g},v_{g}, w_{g}$ essentially they have no physical meaning which means that the smaller the value of $n$ in the summation of Eq. \ref{Error_fun} the larger the reliability of method.\\
The normalization of these wave functions is defined as
\begin{equation}
N = \int_{0}^{\infty}[g^{2}(r)+f^{2}(r)]dr = 1.
\end{equation} 
Other authors use alternatively computational techniques based on expansions of the corresponding wave function into classical polynomials such as Laguerre, Legendre etc. \cite{Polyin1, Polyin2, Polyin3}. The proposed method, however, utilizes Neural Networks that can easily approximate every type of function and provide accurate solutions compared to other methods\cite{Hornik}. Moreover, they are reliable even in environments with error fluctuations.  \\

After the above, in order to find $f(r)$ and $g(r)$, one has to minimize an 
error function ${\cal F} ({\bf u}, {\bf v}, {\bf w})$ of the form 
\cite{Pap-Lag-1,Pap-Lag-2}
\begin{eqnarray}
\label{Error_fun}
{\cal F} = N^{-1}  \sum_{i=1}^{n} \left\{
 \left[ \frac{df(r_{i})}{dr}+\frac{f(r_{i})}{r_{i}} - \frac{m c^{2}-E+V(r_{i})}{\hbar c}g(r_{i})\right]^{2} \right.
\nonumber \\  
 \left. + \left[\frac{dg(r_{i})}{dr}-\frac{g(r_{i})}{r_{i}} - \frac{m c^{2}+E-V(r_{i})}{\hbar c}f(r_{i})\right]^{2} 
 \right\}.  
\end{eqnarray}
The minimization provides the values of the set of parameters ${\bf u}, {\bf v}, {\bf w}$ and 
the binding energy $\epsilon_b$ of the lepton in the Coulomb field of the nucleus 
\cite{Sens,Kenneth} given by
\begin{equation}
\label{binding_en}
\epsilon_b = E - m c^2
\end{equation}
where the total energy of the lepton $E$ is given by
\begin{equation}
E=\frac{m c^{2}\int_{0}^{\infty}[g^{2}(r)+f^{2}(r)]dr+\int_{0}^{\infty}V(r)[g^{2}(r)-f^{2}(r)]dr}{\int_{0}^{\infty}[g^{2}(r)-f^{2}(r)]dr} \, .
\end{equation}

In the present work, we have derived efficient algorithms in order to solve the above quantum 
mechanical eigenvalue problem through the optimization of ${\cal F}$
(for a great number of $n$ in Eq. (\ref{Error_fun})) as presented below.

\section{Description of the minimization methods}

The proposed software, utilizes three different optimization methods: 
a genetic algorithm, a particle swarm optimization method and a simulated annealing 
method, to solve the Dirac equation. The binding energy $\epsilon_b$ of the Dirac 
particle (lepton) orbiting a chosen atomic orbit is determined from the minimum energy, 
$E_{min}=\epsilon_b$, satisfying the eigenvalue problem, Eqs. (\ref{1_Dir_Eq_fr}) and (\ref{2_Dir_Eq_gr}).
In the following each, of these methods is described in detail.

\subsection{Genetic algorithm}

Genetic algorithms are methods based on the so called genetic operations of natural 
selection, reproduction and mutation. They have been used successfully in many areas 
such as combinatorial problems \cite{giannis-key1}, neural network training 
\cite{giannis-key2,giannis-key3}, electromagnetic \cite{giannis-key4}, 
design of water distribution networks \cite{giannis-key5} etc. 
The main steps of the used genetic algorithm are shortly described
in \cite{giannis-key6} and they have:
\begin{itemize}
\item Step 1 \textbf{(initialization}): 

\begin{itemize}
\item Generate $N$ uniformly distributed random points (chromosomes) and
store them to the set $S$.
\item Set iter=0 
\item Set $p_{s}$ the selection rate
\item Set $p_{m}$ the mutation rate
\item Set IMAX the maximum number of allowed iterations.
\end{itemize}
\item Step 2 \textbf{(evaluation)}: Evaluate the function value of each
chromosome. 
\item Step 3 \textbf{(termination check)}: If termination criteria are hold
terminate. The termination criteria of the used algorithm are based
on asymptotic considerations. At every generation denoted by iter,
the variance $\sigma^{(\mbox{iter})}$ of the best located value is
recorded. If there is not any improvement for a number of generations,
it is highly possible that the global minimum is already found and
hence the algorithm should terminate. Also if iter $>=$ IMAX then terminate.
\item Step 4 \textbf{(genetic operations)}: 

\begin{itemize}
\item \textbf{Selection}: Select $m\le N$ parents from $S$. The selection
is performed using the tournament selection technique. 
\item \textbf{Crossover}: Create $m$ new points (offsprings) from the previously
selected parents. 
\item \textbf{Mutation}: Mutate the offsprings produced in the crossover
step with probability $p_{m}$. 
\end{itemize}
\item Step 5 \textbf{(replacement)}: Replace the $m$ worst chromosomes
in the population with the previously generated offsprings. 

\item Step 6$ $

\begin{itemize}
\item \textbf{Set }iter=iter+1 
\item \textbf{goto} step 2 
\end{itemize}
\end{itemize}

\subsection{Particle swarm optimization }

The swarm optimization algorithm (PSO) was initially suggested
by Kennedy and Eberhart \cite{kennedy} and it is an evolutionary 
type algorithm based on population of candidate solutions (swarm 
of particles) which move in an n-dimensional search space. Every 
particle $i$ is assigned its current position ${\bf x}_i$ and the 
corresponding velocity ${\bf u}_i$. These two vectors are repeatedly 
updated, until a predefined convergence criterion is met. 
The PSO method has been applied to a wide range of problems
\cite{pso1,pso2,pso3,pso4} and here we use a variant of the original 
method described in \cite{pso_amc}. The main steps are:
\begin{enumerate}
\item \textbf{Initialization}. 

\begin{enumerate}
\item \textbf{Set} $k=1$ (iteration counter). 
\item \textbf{Set }the number of particles $m$. 
\item Set the maximum number of iterations $k_{\mbox{max}}$
\item \textbf{Initialize} randomly the positions of the $m$ particles $x_{1},x_{2},\ldots,x_{m}$ 
\item \textbf{Initialize} randomly the velocities of the $m$ particles
$u_{1},u_{2},\ldots,u_{m}$ 
\item \textbf{For} $i=1,\ldots,m$ \textbf{ do} $p_{i}=x_{i}$. 
\item \textbf{Set} $p_{\mbox{best}}=\arg\min_{i\in1,\ldots,m}f\left(x_{i}\right)$ 
\item \textbf{Set} $p_{\mbox{old}}=p_{\mbox{best}}$ 
\end{enumerate}
\item \textbf{For} i = 1, . . ., m \textbf{Do}

\begin{enumerate}
\item \textbf{Update} the velocity $u_{i}$ with the equation
\begin{equation}
 u_i=\psi_{1}r_{1}\left(p_{i}-x_{i}\right)+\psi_{2}r_{2}\left(p_{\mbox{best}}-x_{i}\right)
\end{equation}
where $\psi_{1}$ and $\psi_{2}$ are positive constants.
\item \textbf{Update} the position $x_{i}$ as
and $p_{\mbox{best}}$ 
  \begin{equation}
    x_i=x_u+u_i
  \end{equation}

\item \textbf{Calculate} the objective function for particle $i$, $f\left(x_{i}\right)$ 
\item \textbf{If} $f\left(x_{i}\right)\le f\left(p_{i}\right)$ \textbf{then}
$p_{i}=x_{i}$ 
\end{enumerate}
\item \textbf{End For }
\item \textbf{Termination} Check Step 

\begin{enumerate}
\item \textbf{Set} $p_{\mbox{best}}=\arg\min_{i\in1,\ldots,m}f\left(x_{i}\right)$
\item \textbf{Calculate} the variance $\sigma^{(k)}$ of the best located
value (as in Genetic Algorithm case)
\item \textbf{If} there is not any improvement for a number of generations,
terminate and report $p_{\mbox{best}}$ as the discovered minimum.
\end{enumerate}
\item \textbf{Set} $k=k+1$. \textbf{If} $k\ge k_{\mbox{max}}$ \textbf{then
terminate}.
\item \textbf{Goto} Step 2. 
\end{enumerate}

\subsection{Simulated Annealing}

Simulated annealing \cite{simulated} mimics the annealing process
to solve an optimization problem. The temperature parameter $T$ controls
the search which typically starts off high and is slowly \textquotedbl{}cooled\textquotedbl{}
or lowered in every iteration. At each iteration a series of new points
are generated. If the new point has a better function value it replaces
the current point and iteration counter is incremented. It is possible,
however, to accept and move forward with a worse point, but the probability
of doing so is directly dependent on the temperature $T$ (this step,
sometimes, helps identify a new search region in hope of finding a
better minimum and protects the algorithm from being trapped in local
minima).

A typical description of SA algorithm is as follows \cite{Ody-Sim_ann,Ody-Vla_CAM,Ody-Ley_ACM}: 
\begin{enumerate}
\item \textbf{Set} $k=0$, $T_{0}>0$. \textbf{Sample} $x_{0}$ as the initial
point. 
\item \textbf{Set} $N_{eps\mbox{}}>0$, a positive integer
\item Set $\epsilon>0$, a small positive double precision value.
\item Set $r_{T}>0,\ r_{T}<1$, a positive double precision value.
\item For $i=1,\ldots N_{\mbox{eps}}$ 

\begin{enumerate}
\item \textbf{Sample} a point y
\item \textbf{If $f(y)\le f\left(x_{k}\right)$ then $x_{k+1}=y$ }
\item \textbf{Else Set} $x_{k+1}=y$ with probability $\min\left\{ 1,\exp\left(-\frac{f(y)-f\left(x_{k}\right)}{T_{k}}\right)\right\} $
\end{enumerate}
\item \textbf{EndFor}
\item \textbf{Set} $T_{k+1}=T_{k}r_{T}$
\item \textbf{Set} $k=k+1$. 
\item If $T_{k}\le\epsilon$ terminate
\item \textbf{Goto} step 2. 
\end{enumerate}

\section{ Software documentation }

Although this method is complete, it requires many calculation steps for each 
matrix element. In order to verify the method, we have used it for calculating 
the ground state of atomic and nuclear systems by solving the Dirac equations. 
We have solved these problems 
analytically for any excitation and for an arbitrary value of the total angular 
momentum. These solutions are the only examples of general analytic solution of 
the Dirac equation in three dimensional space for the study of nuclear systems 
composed of N nucleons.

In order to carry out our algorithm systematically,
we choose some maximum value $k_{max}$ for the maximum number of iterations allowed \cite{pso_amc}. A computer code implementing 
the formalism described above for one row states was developed. The code 
starts from one coordinate states with a given value for $k_{max}$.

\subsection{Installation }

The user should issue the following commands to build the program
Dirac Solver (under some Unix machine):
\begin{enumerate}
\item Download the code DiracSolver.tar.gz
\item gunzip DiracSolver.tar.gz
\item tar xfv DiracSolver.tar
\item cd DiracSolver
\item qmake .
\item make
\end{enumerate}
The main executable is called DiracSolver 

\subsection{User interface }

The graphical interface of the program is written entirely in Qt and
it contains the following tabs:
\begin{enumerate}
\item \textbf{Settings tab}. A screenshot of this tab is show in Fig \ref{fig:figure1}.
The user can select the desired global optimization method (Genetic
algorithm, Pso or Simulated Annealing) and subsequently he can alter
some default parameters of the selected method such as the number
of chromosomes in genetic algorithm, the initial temperature in Simulated
Annealing etc. Also, under the section Equation Settings the user can change the number of hidden nodes
used by the neural network for the Dirac equation and he can select the desired material for the equation. 
This list contains 9 materials.
\item \textbf{Run tab}. A typical screenshot of this tab is outlined in
Fig \ref{fig:figure2}. The user can start or terminate the optimization
process through two buttons located here.
\item \textbf{Graph tab}. The final outcome of the procedure is displayed
in this tab. A typical screenshot is demonstrated in Fig \ref{fig:figure3}. The user can save
the plot for the material either in png format or in text format ideal for programs such as Gnuplot.
Also in this tab the user can save the parameters of the optimization method as well as the parameters 
of the neural network in a text file using the button entitled SAVE PARAMETERS.
\end{enumerate}

\section{Application of the method in a physical system}

In this section, we apply the new algorithms to find the ground state (1s state) wave 
functions $f(r)$ and $g(r)$ of Eqs. (\ref{1_Dir_Eq_fr}) and (\ref{2_Dir_Eq_gr}), i.e.
large and small component, respectively, of the Dirac 
equations (\ref{eq:frequation}) and (\ref{eq:grequation}), that describe the motion of 
a muon bound in the Coulomb field created by the $^{208}Pb$ nucleus, a system known
as "the mu-mesic $^{208}Pb$ atom". We stress that the application of the new methods, 
to find the basis (or excited) states Dirac wave functions of other quantum systems 
like  hydrogenic-type electron-nucleus systems, the tau-mesic atoms \cite{Oset-91}, etc., is 
straightforward. 

The potential energy $V({\bf r})$ entering Eqs. (\ref{1_Dir_Eq_fr}) and (\ref{2_Dir_Eq_gr}),
for the nuclear Coulomb field originating from an extended nuclear charge density distribution $\rho({\bf r})$, 
is calculated by \cite{Kosm-Lag}
\begin{equation}
V({\bf r})=-e^{2}\int \frac{\rho({\bf r}^{\prime})}{|{\bf r}-{\bf r}^{\prime}|}
d^3 {\bf r}^{\prime}.
\label{Coulomb-Poten}
\end{equation}
The nuclear charge density $\rho({\bf r})$ in this work is obtained from model 
independent analysis of the electron scattering experimental data of Ref. \cite{deVries}.
For the chosen nuclear systems (we assume their charge distribution is
spherically symmetric)
the radial charge density distribution entering Eq. 
(\ref{Coulomb-Poten}) is described by a two-parameter Fermi distributions 
\cite{deVries,Shanker-79} of the form
\begin{equation}
\rho(r)=\frac{\rho_{0}}{1+e^{(r-c)/z}} \, .
\label{Fermi-distr}
\end{equation}
For the parameters $c$ and $z$ (known as radius and thickness parameter, respectively),
in the case of $^{208}Pb$ nucleus, we adopted the values of Ref. \cite{deVries}.
The parameter $\rho_{0}$ is determined from the normalization of $\rho({\bf r})$
\cite{deVries}.
The root mean square charge radius of this nuclear isotope (representing the extension 
of the nuclear finite size), is $R_{ms} \approx 5.5$ fm.
It should be noted that, for both the Schroedinger and Dirac solutions 
one has to take into account some (may be significant) corrections to the
potential $V(r)$ such as the nuclear polarization \cite{Kosm-Lag}, the 
vacuum polarization \cite{Kitano,Sens}, etc. In the present work, we have taken into account an effective potential $V_{vp}$ due to
the vacuum polarization corrections described by an effective potential 
as in Refs. \cite{Pap-Lag-1,Kosm-Lag}.

We furthermore note that, in other similar works (e.g. Ref. \cite{Stoica-AHEP,Stoica-PRC}), authors employ a phenomenological description for the $V(r)$ 
corresponding to a uniform distribution of the nuclear charge within the mean charge 
radius $R_{ms}=R_c$ of the form
$$
V(r)=\begin{cases}
\frac{(Z-1)e^2}{2R_{c}}\left[3-\left(\frac{r}{R_c}\right)^2\right] &, 
\qquad r\le R_c,\\ \frac{(Z-1)e^2}{r} &, \qquad r> R_c \, .
\end{cases} 
$$
It must be mentioned that, for large distances the 
matching of the wave functions $f(r)$ and $g(r)$ to their asymptotic behavior 
was done as in \cite{Shanker-79,Sens,Kenneth}. For light nuclei this matching is achieved at 
$r \approx 60 - 70$ fm and for heavy and very heavy nuclear isotopes at 
$r \approx 40 - 50$ fm.

\begin{figure}
\includegraphics[scale=0.5]{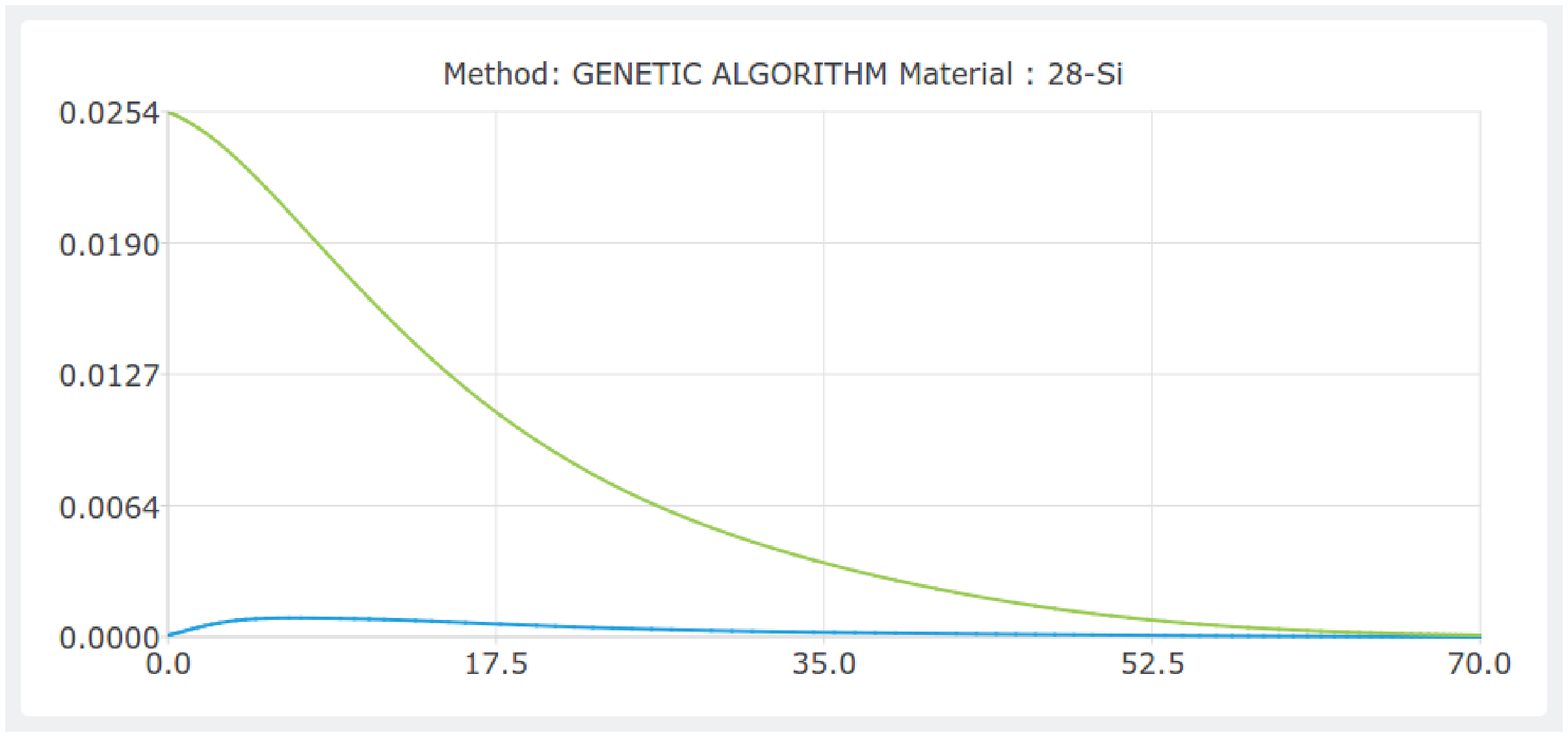} 
\caption{The output for material ${}^{28}\mbox{Si}$ using 
Genetic Algorithm.}
\label{genetic_output}
\end{figure}

\begin{figure}
\includegraphics[scale=0.5]{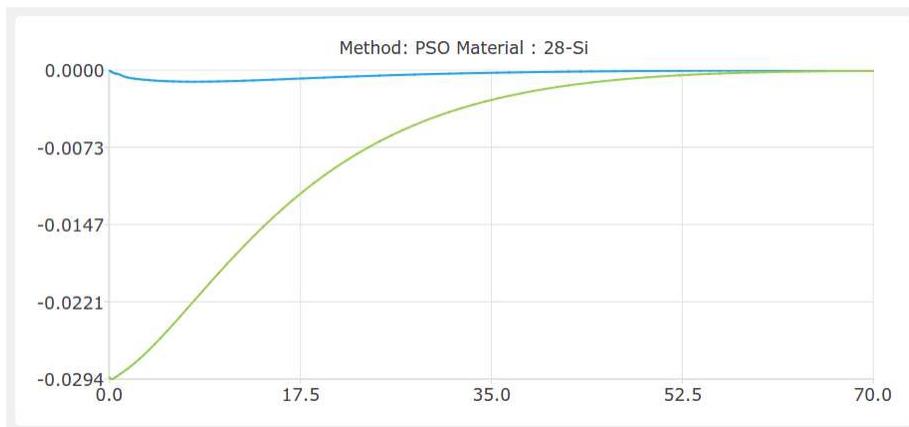} 
\caption{The output for material ${}^{28}\mbox{Si}$ using
Particle Swarm Optimization.}
\label{pso_output}
\end{figure}

\begin{figure}
\includegraphics[scale=0.5]{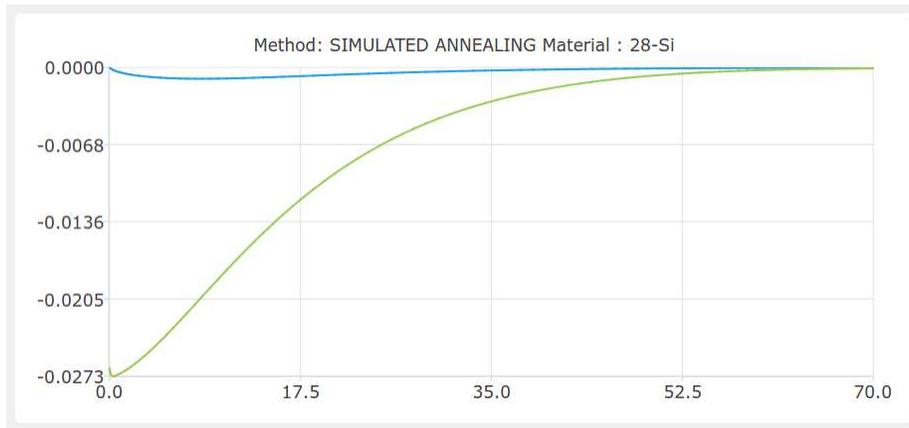}
\caption{The output for material ${}^{28}\mbox{Si}$ using 
Simulated Annealing}
\label{sa_output}
\end{figure}

\begin{figure}
\caption{The tab Settings of the application. The user can choose the optimization method and the running material.
\label{fig:figure1}}
\centering{}\includegraphics[scale=0.4]{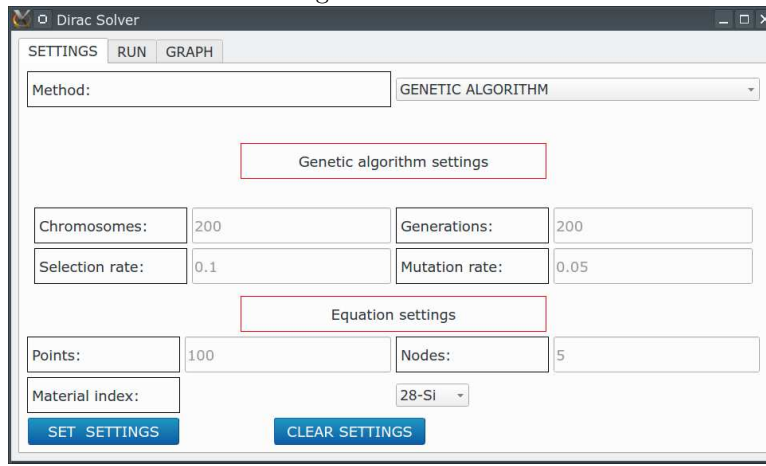}
\end{figure}

\begin{figure}
\caption{The tab Run of the application. The user can initiate the execution of the optimization method or
he can terminate it. 
\label{fig:figure2}}
\centering{}\includegraphics[scale=0.4]{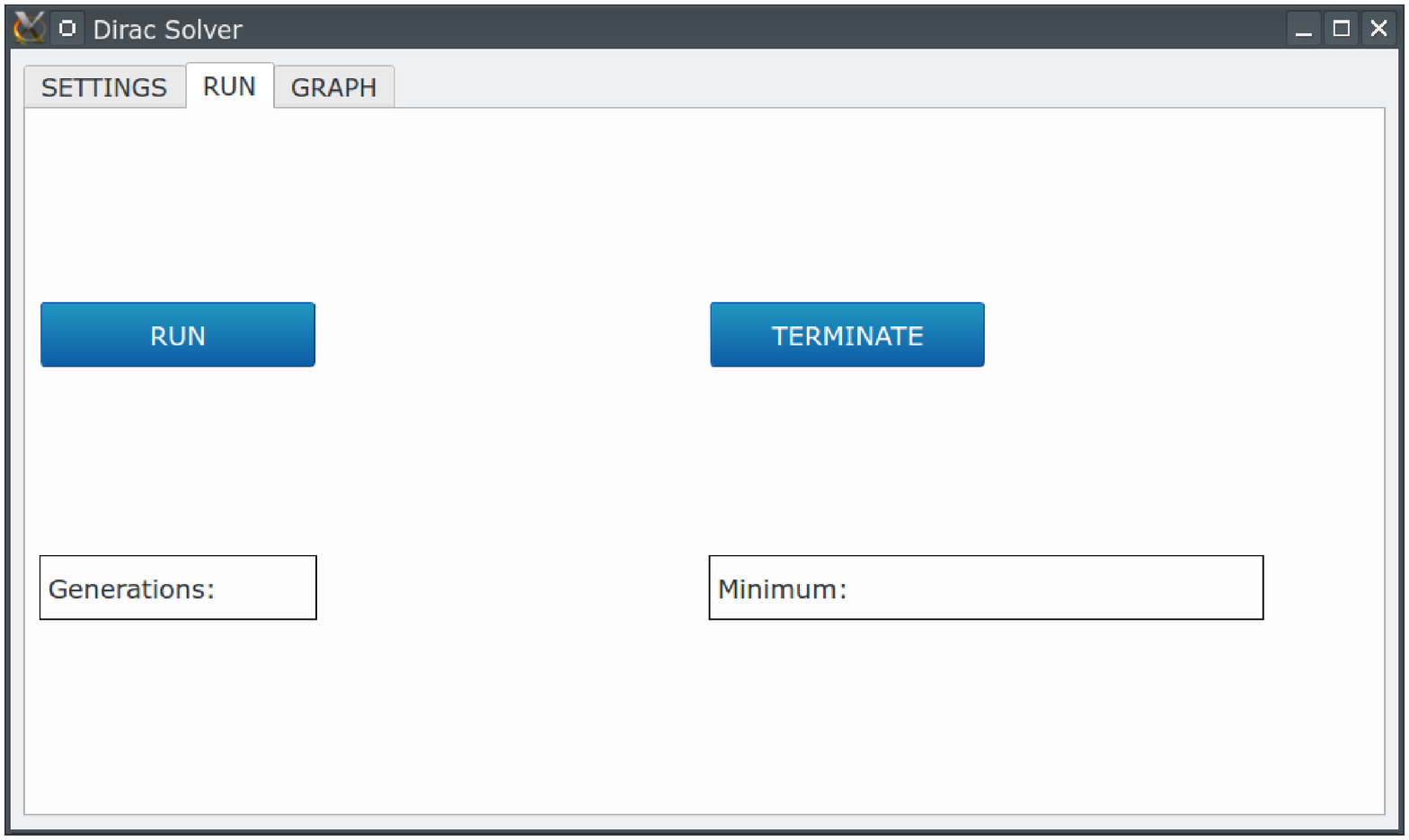}
\end{figure}

\begin{figure}
\caption{The tab Graph of the application. The user can save the plot in png format or he can 
save the plot in column format ideal for programs such as gnuplot or he can save the parameters of the execution.
\label{fig:figure3}}
\centering{}\includegraphics[scale=0.4]{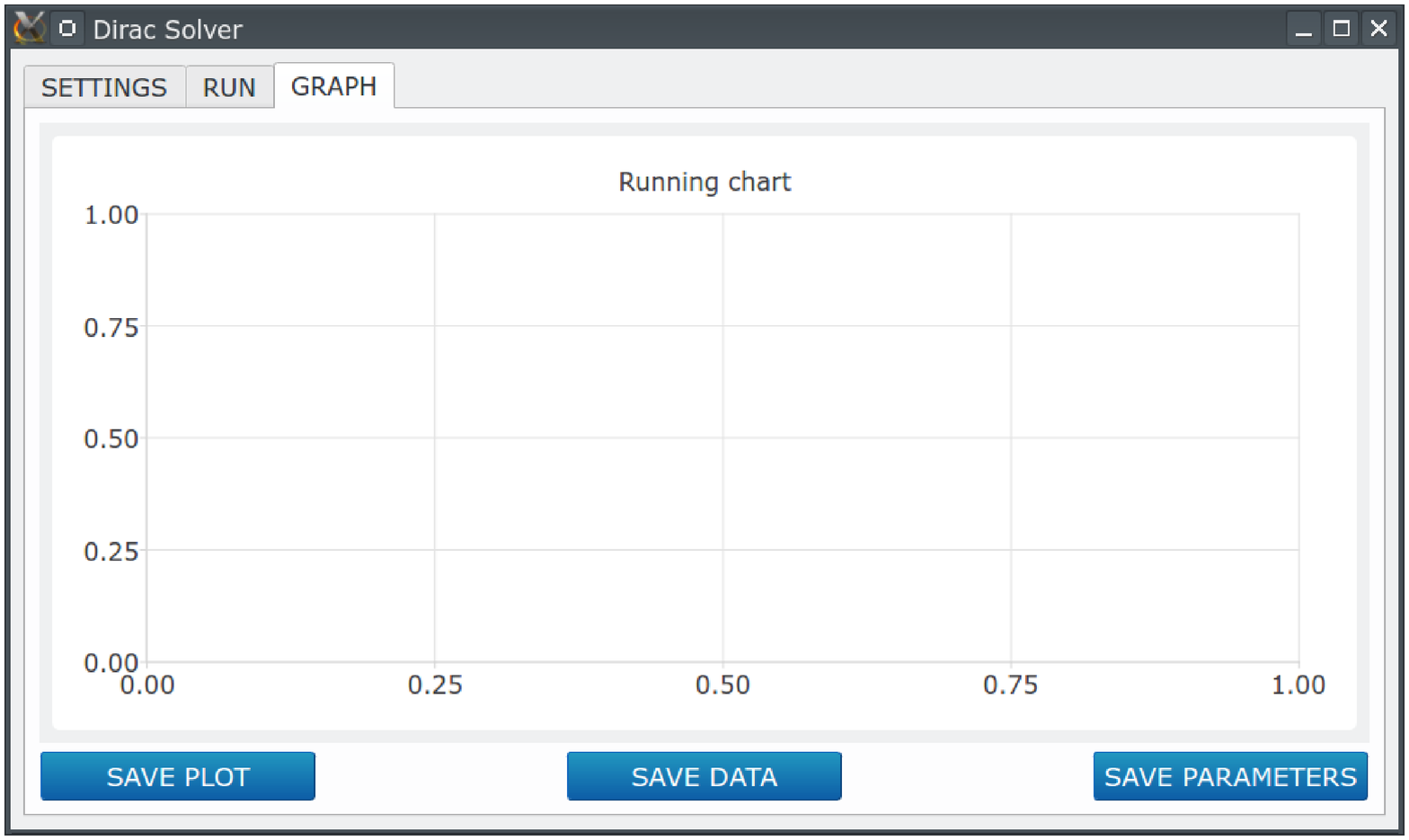}
\end{figure}

The default values for the parameters of the Genetic algorithm, the PSO method 
and the Simulated Annealing method are displayed in Table \ref{tab:algorithms_params}. 
A typical outcome for the three optimization procedures is shown
in Figures \ref{genetic_output}, \ref{pso_output} and \ref{sa_output}. 
Example values obtained through the optimization techniques examined here, for the muon binding energy of Eq. (\ref{binding_en}) are 
listed in Table \ref{tab:energyvalues}. For comparison with the experimental values of $\varepsilon_b$ the reader is referred to Refs.  \cite{Kosm-Lag,Kitano}.

We should mention that, in realistic 
nuclear structure calculations, integrals of the form of Eq. (11) of Ref. \cite{Kosm-Lag} 
(see also Ref. \cite{Giannak-Kosm}) may be explicitly obtained without the need to utilize
approximation adopting an average muon wave function $\langle\Phi_{\mu}\rangle$ at $r = 0$. This, 
of course, implies that the evaluation of the required nuclear matrix elements becomes, in general, 
more complicated and time consuming. However, the obtained analytic-type expressions for the solution of 
the Dirac wave functions given by Eqs. (\ref{eq:frequation}) and (\ref{eq:grequation})
facilitates the calculations of the necessary physical observables.

Before closing this section it is worth making the following remarks: the advantages
of utilizing  Eqs. (\ref{eq:frequation}) and (\ref{eq:grequation}) for the muon wave 
functions become more evident when one calculates the incoherent rate in a muonic process 
like e.g in the ordinary muon capture, the muon to electron conversion ($\mu^- \to e^-$), the muon to positron conversion ($\mu^- \to e^+$), etc., where one has to face a large number 
of (double or multiple) numerical integrations corresponding to transition integrals from 
the initial (ground) to a final state of the chosen system. In addition, due to the fact 
that in our method the muon wave functions are available in a convenient way, i.e. as linear 
combinations of well-behaved (sigmoid) functions, the use of extrapolation and/or 
interpolation techniques required in other methods  \cite{Giannak-Kosm} is avoided and the accuracy 
of the results obtained this way is very high.


\section{Summary and Conclusions}

In many physical problems, one needs to describe accurately the motion
of bound leptons (e, $\mu$, $\tau$, etc.) in atomic orbits. Because this
motion is relativistic, one must go beyond the usual calculation of the 
lepton density at the site of the atomic nucleus obtained through the use of
the Schroedinger wave function. This means that one should solve numerically 
the Dirac equation in the field of the finite size nuclei.  
Specifically, for high Z values of the atomic nuclei, the motion of a 
muon, tau leptons, etc. is relativistic, and the small component $g(r)$ of 
their wave functions is non-negligible. Therefore, for such calculations 
one requires the additional contribution associated with that component 
of the Dirac equation.

In this paper, we have constructed three independent algorithms, based 
on three different stochastic optimization methods, for solving 
numerically the Dirac equation. These methods used a relativistic
formalism to evaluate the radial part of these wave functions. 

We applied the algorithm to find the ground state Dirac wave functions
of a lepton-nucleus 
system, however, it is straightforward to apply these algorithms to 
obtain the basis 
states of other quantum systems like the tau mesic atoms. Also, they could
be easily used for finding the wave functions that describe states of 
lepton-nucleus systems in which the lepton is orbiting in higher states
(excited states).

The comparison of the algorithms presented above with the state of the art 
symmetrization algorithms, proves the computational advantages of our present 
algorithms. 

\section{Acknowledgements}
Dr.~Odysseas Kosmas wishes to acknowledge the support of EPSRC, United Kingdom, via grand EP/N026136/1 "Geometric Mechanics of Solids".
 
\section{Appendix A}

The general form of the 3-dimensional relativistic bound-state wave 
function $\psi$ describing the motion of a Dirac particle in central force 
system is written as
\cite{Rose}
\begin{equation}
E\psi=\left[-i\gamma_{5}\sigma_{r}\left(\frac{\partial}{\partial r}+\frac{1}{r}-\frac{1}{r}\beta K\right)+V(r)+m_{i}\beta\right]\psi
\label{3D-Dirac-Eq}
\end{equation}
where $\gamma_{5}$, $\beta$, $\sigma_{r}$ and $K$ are the (4x4) matrices
\begin{eqnarray}
\gamma_{5} & = & \left(\begin{array}{cc}
0 & 1\\
1 & 0
\end{array}\right), \qquad
\beta=\left(\begin{array}{cc}
1 & 0\\
0 & -1
\end{array}\right)\label{eq:eq2}
\end{eqnarray}
\begin{eqnarray}
\sigma_{r} & = & 
\left(\begin{array}{cc}
\sigma\cdot {\bf r} & 0\\
0 & \sigma\cdot {\bf r}
\end{array}\right), \qquad 
K=\left(\begin{array}{cc}
\sigma\cdot \ell+1 & 0\\
0 & -(\sigma\cdot \ell+1)
\end{array}\right)\label{eq:eq3}
\end{eqnarray}
(in the above matrices the unit stands for the unit matrix) 
where ${\boldmath \sigma} = (\sigma_1, \sigma_2, \sigma_3)$, with 
$\sigma_i, i=1,2,3$, being the well known Pauli matrices \cite{Feschach}. 
The symbols $E$ and $V(r)$ denote the total energy and potential energy,
respectively, and $\bf \ell$ represents the orbital angular momentum of the 
Dirac particle. The latter operator is defined by ${\bf \ell} = \imath {\bf r}\times {\bf \nabla}$. 
Also, $m_i$ is the reduced mass of the lepton (e, $\mu$, $\tau$), etc. 
given in Eq. (\ref{eq:red-mass}). 

Even though $\psi$ is not an eigenstate of ${\bf \ell}^2$, however, 
the upper and lower components of $\psi$, $\psi_\alpha$ and $\psi_\beta$ 
respectively, are separately eigenstates (they have fixed total
angular momentum $j$ 
and spin $s$ quantum numbers), so that $\psi$ is written as 
\begin{equation}
\psi = \left(\begin{array}{c}
\psi_{\alpha} \\
\psi_{\beta}
\end{array}\right).
\end{equation}
Then, the solutions of the 3-Dimensional Dirac equation (\ref{3D-Dirac-Eq}),
describing a Dirac particle in a central force field, in spherical 
coordinates take the form \cite{Kitano}
\begin{equation}
\psi = \psi^{\mu}_{\kappa} = \frac{1}{r}
\left(\begin{array}{c}
 f(r)_{\kappa} \Phi_{\kappa \mu} (\theta ,\phi) \\
\imath g(r)_{\kappa} \Phi_{-\kappa \mu} (\theta ,\phi) \, 
\end{array}\right).
\label{Dirac-WF}
\end{equation}
The quantum numbers $\kappa$ and $\mu$ describing the above wave functions,   
are eigenvalues of the operators $-K$ and the projection ${\bf j}_z$ of the total
angular momentum operator ${\bf j} = {\bf \ell} + {\bf s}$  (${\bf s}$
denotes the spin operator of the Dirac particle for which we assume $s=1/2$)
\cite{Kitano}. 

In Eq. (\ref{Dirac-WF}), the radial wave functions $g(r)$ and 
$f(r)$ are determined through the two coupled first-order differential equations,
Eqs. (\ref{1_Dir_Eq_fr}) and (\ref{2_Dir_Eq_gr}), and 
$\Phi_{\kappa \mu} (\theta ,\phi)$
is a spherical spinor in the direction $(\theta ,\phi)$
(the known spin spherical harmonics) defined by the equation
\begin{equation}
\Phi_{\kappa \mu} = \sum_{m_\ell m_s} \langle \ell m_\ell \frac{1}{2} m_s
\vert (\ell \frac{1}{2}) j m \rangle 
Y_{\ell m_\ell} (\theta , \phi) \chi_{\frac{1}{2}m_s}
\label{spin-spherical}
\end{equation}
($Y_{\ell m_\ell}$ represent a spherical harmonic) where $j= \vert\kappa\vert - \frac{1}{2}$ which gives 
$\kappa = \mp(j+1/2)$ with $j = \ell \pm 1/2$. 
The symbol $\langle\vert\rangle$ 
in Eq. (\ref{spin-spherical}) denotes the well known 
Clebsch-Gordan coefficients which are numbers arising in angular 
momentum couplings in quantum mechanics \cite{Feschach}. 

\bibliographystyle{unsrt}

\newpage

\begin{table}
\caption{The default parameters for the Algorithms used. \label{tab:algorithms_params}}
\centering{}
\begin{tabular}{|c|c||c|c||c|c|}
\hline
\hline
\multicolumn{2}{|c||}{Genetic Algorithm}&\multicolumn{2}{|c|}{PSO method}& 
\multicolumn{2}{|c||}{Simulated Annealing} \\
\hline 
PARAMETER & VALUE & PARAMETER & VALUE & PARAMETER & VALUE \\
\hline
\hline  
N     & 200  &       $m$        & 200 & $T_{0}$          & 80.0 \\
IMAX  & 200  & $k_{\mbox{max}}$ & 200 & $N_{\mbox{eps}}$ & 200 \\
$p_s$ & 0.1  & $\psi_1$         & 1.0 & $\epsilon$       & $8\times10^{-5}$ \\
$p_m$ & 0.05 & $\psi_2$         & 1.0 & $r_{T}$          & 0.9  \\
\hline 
\hline 
\end{tabular}
\end{table}

\begin{table}
\caption{Muon binding energies obtained for each of the algorithmic methods for a set of interesting 
muonic atoms (see the text).
\label{tab:energyvalues}}
\centering{}
\begin{tabular}{|c|c|c|c|}
\hline 
  &&& \\
MATERIAL & GENETIC & PSO & ANNEALING \\
\hline
\hline  
&  &  & \\
${}^{28}\mbox{Si}$  & 0.604  & 0.581  & 0.575  \\
${}^{48}\mbox{Ti}$  & 1.585  & 1.510  & 1.504  \\
${}^{64}\mbox{Zn}$  & 2.443& 2.490    & 2.415  \\
${}^{98}\mbox{Mo}$  & 4.093 & 4.089   & 4.059  \\
${}^{124}\mbox{Sn}$ & 5.319 & 5.319   & 5.318  \\
${}^{156}\mbox{Gd}$ & 7.625 & 7.627   & 7.624  \\
${}^{186}\mbox{W}$  & 9.222 & 9.224   & 9.210  \\
${}^{208}\mbox{Pd}$ & 10.599 & 10.599 & 10.545 \\
${}^{238}\mbox{U}$  &12.363 & 12.363  & 12.433 \\

\hline 
\hline 
\end{tabular}
\end{table}

\section{Appendix B: PROGRAM SUMMARY}

\textit{Title of program}: DiracSolver

\textit{Catalogue identifier}:

\begin{flushleft}
\textit{Program available from}: CPC Program Library, Queen's University
of Belfast, N. Ireland.
\par\end{flushleft}

\begin{flushleft}
\textit{Computer for which the program is designed and others on which
it has been tested}: The tool has been tested on Linux, Android and
Windows. The tool is designed to be portable in all systems running
the GNU C++ compiler using the QT programming library.
\par\end{flushleft}

\emph{Licensing provisions}: GPLv3

\begin{flushleft}
\emph{Installation}: Technological Educational Institute of Epirus,
Greece.
\par\end{flushleft}

%

\textit{Programming language used}: GNU-C++\emph{}\\
\emph{Memory required to execute with typical data}: 200KB.\textit{}\\
\textit{No. of bits in a word}: 64\emph{}\\
\emph{No. of processors used}: many\emph{}\\
\emph{Has the code been vectorized or parallelized?}: No.\emph{}\\
\emph{No. of bytes in distributed program,including test data etc}.:
100 Kbytes.\emph{}\\
\emph{Distribution format}: gzipped tar file.\emph{}\\
\emph{Keywords}: Global optimization, stochastic methods, genetic
algorithms\emph{}\\
\emph{Nature of physical problem}\textbf{: }.\emph{}\\
\emph{Solution method}: .\emph{}\\
\emph{Typical running time}: Depending on number of processing nodes and 
the used optimization function.
\emph{Nature of the problem}:The software tackle the problem of solving the Dirac equation using stochastic optimization methods.\\
\emph{Solution method}:The software utilizes three stochastic optimization methods for the solution of Dirac equation in the form of neural networks. 
The used methods are: genetic algorithm, particle swarm optimization and simmulated annealing.\\


\end{document}